\documentclass[12pt, a4paper]{jpconf} %amssymb,amsmath
\usepackage{citesort}
\usepackage{graphicx}

\begin{document}

\title{Spatial properties of $\pi-\pi$ conjugated network in semicrystalline polymer thin films studied by intensity x-ray cross-correlation functions}

\author{R P Kurta$^1$,  L Grodd$^2$, E Mikayelyan$^2$, O~Y~Gorobtsov$^1$$^,$$^3$, I Fratoddi$^4$, I Venditti$^5$, M Sprung$^1$, S Grigorian$^2$  and I A Vartanyants$^1$$^,$$^6$}

\address{$^1$ Deutsches Elektronen-Synchrotron DESY, Notkestra\ss e 85, D-22607 Hamburg, Germany}

\address{$^2$ Department of Physics, University of Siegen, Walter-Flex-Strasse 3, D-57072 Siegen, Germany}

\address{$^3$ National Research Center 'Kurchatov Institute', Kurchatov Square 1, 123182 Moscow, Russia}

\address{$^4$ Department of Chemistry and Center for Nanotechnology for Engineering (CNIS), University of Rome Sapienza, P.le A. Moro 5, I-00185 Rome, Italy}

\address{$^5$ Department of Chemistry, University of Rome Sapienza, P.le A. Moro 5, I-00185 Rome, Italy}

\address{$^6$ National Research Nuclear University, ``MEPhI'', 115409 Moscow, Russia}

\eads{\mailto{ruslan.kurta@desy.de}, \mailto{grigorian@physik.uni-siegen.de}}

\begin{abstract}
We present results of x-ray study of spatial properties of $\pi-\pi$ conjugated networks in polymer thin films. 
We applied the x-ray cross-correlation analysis to x-ray scattering data from blends of poly(3-hexylthiophene) (P3HT) and gold nanoparticles.
The Fourier spectra of the intensity cross-correlation functions for different films contain non-zero components of orders $n=2,4$ and $6$ measuring
the degree of structural order in the system.

\end{abstract}

\section{Introduction}

Charge carrier mobility is a decisive parameter of organic field-effect
transistors (OFETs). It determines the overall device performance. 
Semicrystalline conjugated polymers are promising candidates for OFETs.
Typical features of these polymers are mixtures of poor and well organized domains. 
Usually the well organized domains are addressed to be crystalline and assumed to strongly improve device performance. 
Among different conjugated polymers polythiophenes have received an increasing attention in recent years due to its attractiveness for  OFETs and solar cell applications \cite{Salleo}. 
One of the most often used polythiophene is poly(3-hexylthiophen) (P3HT)  
which crystalline domain sizes can vary from tens to few hundreds of nanometers, depending on preparation techniques \cite{Prosa,Rahimi}. 

A conventional technique for structural analysis of organic thin films is grazing incidence 
x-ray diffraction (GIXD) which, in many cases of pristine P3HT and its blends, reveals a preferential edge-on orientation \cite{Kim}.  
Generally it is assumed that such orientation with $\pi-\pi$ stacking being parallel to the sample surface enhances OFET performance \cite{Sirring}.
Typically, scattering signal from spin cast polythiophene films probed by GIXD is rather weak.
This is especially true for the in-plane peaks associated with $\pi-\pi$ conjugated network, even measured at highly intense synchrotron sources. 
Moreover, structural information on the scale of individual polythiophene stacks can only be achieved by combination of 
GIXD analysis with complementary high resolution transmission electron microscopy (TEM) technique \cite{Shabi}. 

Novel x-ray cross-correlation analysis (XCCA) technique \cite{Wochner1,Altarelli,AltarelliErr,Kurta1,Kurta2,Kurta3}
is an attractive way of structural characterization of partially disordered samples that can provide complementary information. 
It is a powerful tool to determine spatially-resolved order-disorder variation from coherent diffraction patterns. 
Recently  nanobeam studies in transmission geometry on pristine P3HT films were successfully realized where XCCA unraveled the local structural information of P3HT nanodomains  hidden in the powder average \cite{Gutt}.
The aim of current study is to determine spatially-resolved order-disorder variation of P3HT 
host matrix upon small addition of gold nanoparticles stabilized with fluorene derivatives (AuNPs-SFL)  from the coherent diffraction pattern by XCCA technique. 
Special goal of our work is to understand the structural properties of the $\pi-\pi$ conjugated P3HT network on the
nanoscales which is likely to have a strong effect on the intermolecular charge transport and thus the electronic properties.

\section{Theory}

We consider coherent x-ray scattering from a partially ordered polymer/AuNPs blend. 
Schematic  setup of x-ray transmission geometry is shown in Fig. 1(a) where the strongest intensity ring is associated with $\pi-\pi$ conjugated network. 
One can expand the scattered intensity $I(\mathbf{ q})$ at each momentum transfer vector $\mathbf{q}=(q,\varphi)$ into angular Fourier series as
\begin{equation}
I(q,\varphi)=I_{0}(q)+2\sum\limits_{n=1}^{\infty}I_{n}(q)\cos(n\varphi),\label{Eq:Cchi}
\end{equation}
where $q$ and $\varphi$ are the radial and azimuthal coordinates of the vector $\mathbf{q}$ in polar coordinate system, 
and $I_{n}(q)$ are the Fourier components of intensity $I(q,\varphi)$. 
Depending on the system under investigation and experimental conditions, these complex Fourier components may encode rich
information about the structure of a single particle in a disordered system \cite{Kurta2}, 
degree of bond-orientational order in liquid crystals \cite{Kurta4} or local structure of amorphous system \cite{Wochner1}. 

The angular XCCA enables direct determination of the Fourier components $I_{n}(q)$ from the ensemble
of diffraction patterns measured at different positions on the sample \cite{Altarelli,AltarelliErr,Kurta1}.
For $i$-th diffraction pattern a two-point cross-correlation function (CCF) can be defined as
\begin{equation}
C^{i,i}(q,\Delta) = \left\langle I^{i}(q,\varphi)I^{i}(q,\varphi+\Delta)\right\rangle _{\varphi}, \label{Eq:CCF1}
\end{equation}
where $\Delta$ is the angular coordinate, $\langle \dots \rangle_{\varphi}$
denotes the angular average around the ring of a radius $q$, and $I^{i}(q,\varphi)$ indicate the intensities measured on the same $i$-th diffraction pattern.
The CCF $C^{i,i}(q,\Delta)$ can be analyzed using a Fourier series decomposition \cite{Altarelli,AltarelliErr,Kurta1},
\begin{equation}
C^{i,i}(q,\Delta)=C^{i,i}_{0}(q)+2\sum\limits_{n=1}^{\infty} C^{i,i}_{n}(q)\cos(n\Delta),\label{Eq:Cqn1}
\end{equation}
where $C^{i,i}_{n}(q)$ are the Fourier components of the CCF, 
particularly $C^{i,i}_{0}(q)\equiv \langle I^{i}(q,\varphi)\rangle_{\varphi}^{2}$, and $C^{i,i}_{n}(q)=|I^{i}_{n}(q)|^2$ for $n\neq 0$.

For a disordered system, the Fourier components $C^{i,i}_{n}(q)$ fluctuate from position to position on the sample \cite{Altarelli,AltarelliErr,Kurta1,Kurta2,Kurta3},
therefore averaging of the CCF (or its Fourier spectra)  over a large number $M$ of diffraction patterns is applied to get the statistically averaged result, 
\begin{equation}
\langle C^{i,i}(q,\Delta) \rangle_{M}=1/M\sum_{i=1}^{M}C^{i,i}(q,\Delta),\label{Eq:CCFaver}
\end{equation}

Another two-point CCF $C^{i,j}(q,\Delta)$ can be defined as an angular intensity correlation function between two different diffraction patterns $i$ and $j$,
\begin{equation}
C^{i,j}(q,\Delta) = \left\langle I^{i}(q,\varphi)I^{j}(q,\varphi+\Delta)\right\rangle_{\varphi},\quad i\neq j, \label{Eq:CCF2}
\end{equation}
where $I^{i}(q,\varphi)$ and $I^{j}(q,\varphi)$ indicate the intensities measured on the $i$-th and $j$-th diffraction patterns, respectively.
Similar to $C^{i,i}(q,\Delta)$, this CCF can be expressed in terms of the Fourier components $C^{i,j}_{n}(q)$ and averaged over a set of diffraction patterns,
$\langle C^{i,j}(q,\Delta) \rangle_{M}=1/M\sum_{i,j}^{M}C^{i,j}(q,\Delta)$, where the summation is done over $M$ random pairs of diffraction patterns with $i\neq j$.
Importantly, the ensemble averaged Fourier components $\langle C^{i,j}_{n}(q) \rangle_{M}$ have vanishing values if there is no correlation
between different diffraction patterns. At the same time nonvanishing values of $\langle C^{i,j}_{n}(q) \rangle_{M}$ indicate the presence of correlations between diffraction patterns.
This property of $C^{i,j}(q,\Delta)$ helps to detect contribution to the scattered intensities that can not be attributed to the sample.
Below we analyze the difference Fourier spectra $\langle \widetilde{C}_{n}(q) \rangle_{M}=\langle C^{i,i}_{n}(q) \rangle_{M}-\langle C^{i,j}_{n}(q) \rangle_{M}$. 
These spectra contain structural information intrinsic to the sample.

\section{Experiment}

\begin{figure}[ht]
\centering
\includegraphics[width=0.6\textwidth]{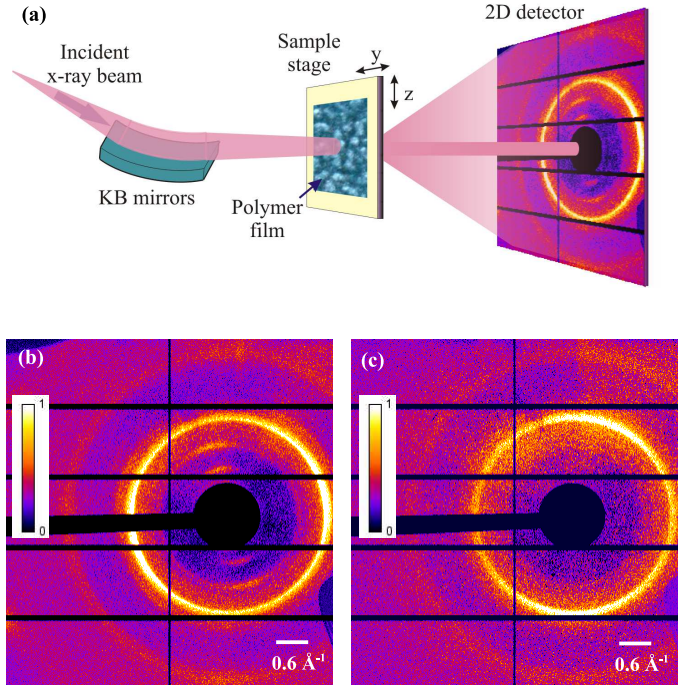}
\caption{(a) Geometry of the scattering experiment
showing the focusing KB mirrors, sample, and 2D detector. (b),(c)
Typical diffraction patterns measured for two different P3HT polymer films with Au nanoparticles.
\label{Fig:ExpGeom}}
\end{figure}

The coherent x-ray scattering experiment was performed at the 
%G{\"o}ttingen instrument for nano-imaging with x-rays 
nanoprobe endstation GINI-X \cite{Kalbfleisch} installed at the coherence beamline P10 of the PETRA III facility at DESY in Hamburg.
The scattering geometry of the experiment is shown in Fig.~\ref{Fig:ExpGeom}(a).
The incident photon energy was chosen to be $13\;\rm{keV}$ and a 2D detector was positioned 
in transmission geometry at $195\;\rm{mm}$ distance from the sample and protected by a beamstop of $34\;\rm{mm}$ in diameter.
The scattering data were recorded on a hybrid-pixel detector Pilatus 1M from Dectris with $981\times1043$ pixels and a pixel size of $172\times 172\;\mu\rm{m}^{2}$.

The sample was mounted on a goniometer, and a film was aligned with its surface perpendicular to the direction of the incident beam.
The beam with a flux of about $5\cdot10^{10}\;\rm{photons/sec}$ was focused on the sample with Kirkpatrick-Baez (KB) mirrors to a spot of about $200\times300\;\rm{nm}^2$ (FWHM).
The sample was scanned in the plane perpendicular to the incident beam direction with a step size larger than the probe size. 
A cryogenic cooling of the film with liquid nitrogen was used during measurements to reduce the sample damage.
The exposure time was chosen to be $0.3\;\rm{sec}$ per image to perform measurement in non-destructive regime. 

P3HT material (molecular weight 44900 g/mol; PDI 1.22 ) has been synthesized at the University of Wuppertal, Germany. 
Gold nanoparticles stabilized with fluorene derivatives (AuNPs-SFL) have been prepared in analogy to literature reports \cite{Fratoddi,Matassa,Quintiliani}.
 A 5~mg/ml blend solution (P3HT:AuNPs-SFL = 1:0.1 by weight) in chloroform were spin cast on rectangular shaped grids with a 15 nm thick $\rm{Si}_{3}\rm{N}_{4}$ membranes (Dune Sciences, Inc.).

\section{Results}
\begin{figure}[ht]
\centering
\includegraphics[width=0.9\textwidth]{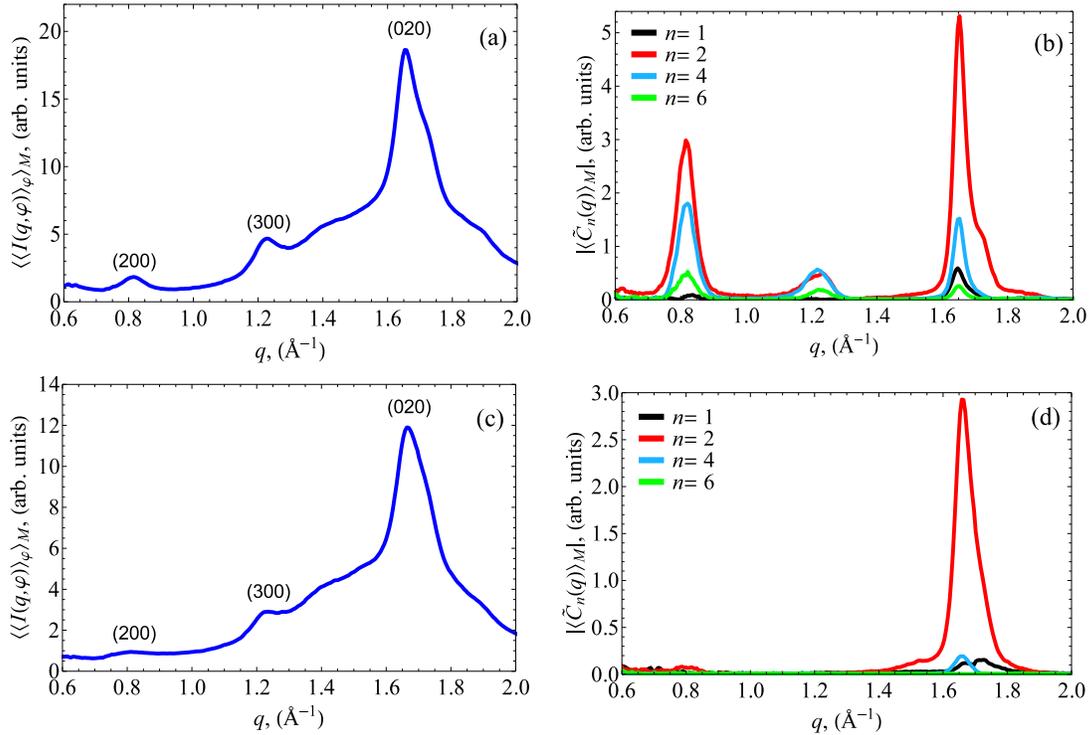}
\caption{(a),(c) Ensemble averaged radial intensity $\langle \langle I(q,\varphi) \rangle_{\varphi}\rangle_{M}$ 
and (b),(d) modulus of the difference spectra $\vert \langle \widetilde{C}_{n}(q) \rangle_{M} \vert$ calculated for two different samples.
\label{Fig:Data1}}
\end{figure}

We measured two datasets for two different samples described above, consisting of $800$ and $1600$ diffraction patterns.
Typical diffraction patterns corrected for the background scattering corresponding to these two samples are shown in Figs.~\ref{Fig:ExpGeom}(b) and \ref{Fig:ExpGeom}(c). 
Radial intensities $\langle \langle I(q,\varphi) \rangle_{\varphi}\rangle_{M}$ averaged
over the full dataset for each sample in the range $0.6 \leq q \leq 2.0\;\mathring{A}^{-1}$ are presented in Figs.~\ref{Fig:Data1}(a) and \ref{Fig:Data1}(c). 
As one can see, both intensity curves look very similar.
The strongest feature present on both diffraction patterns is a $(020)$ ring at $q=1.65\;\mathring{A}^{-1}$, which is related to the $\pi$-stacking conjugated network.
The major difference between two samples is dictated by the orientation of $(020)$ planes and alkyl side chains. 
In the case of mixed orientation [Fig.~\ref{Fig:ExpGeom}(b)] alkyl side chains are tilted with respect to the sample surface, 
producing additional  $(200)$ and $(300)$ peaks at $q=0.8\;\mathring{A}^{-1}$ and $q=1.2\;\mathring{A}^{-1}$, respectively [$(100)$ reflection was not accessible in our experiment because it was covered by the beamstop]. 
In the case of dominating edge-on orientation [Fig.~\ref{Fig:ExpGeom}(c)] alkyl chains are aligned almost perpendicular and $\pi-\pi$ stacking is parallel to the sample surface. 
In this case the $(h00)$ peaks are hardly visible for x-rays in contrast to the $(020)$ planes. 
While intensity distribution around $(020)$ peak has ring-like structure due to randomly distributed domains, the $(200)$ and $(300)$ peaks are strongly anisotropic. 
This is reflected in the two-fold rotational symmetry of the intensity distribution at lower $\mathbf{q}$ [Fig.~\ref{Fig:ExpGeom}(b)]. 
Note, that in the case of preferential edge-on orientation the $(200)$ and $(300)$ peaks are not observed on individual diffraction patterns [Fig.~\ref{Fig:ExpGeom}(c)], but
the ensemble-averaged radial intensity [Fig.~\ref{Fig:Data1}(c)] still shows weak scattering at corresponding $\mathbf{q}$ values.

We performed the cross-correlation analysis of the two measured datasets. 
The resulting difference Fourier spectra $\langle \widetilde{C}_{n}(q) \rangle_{M}$ are presented\footnote{There is also a ring, that partially visible on the detector at higher scattering angles ($q=2.7\;\mathring{A}^{-1}$), which corresponds to scattering from gold nanoparticles. 
Due to a small amount of  AuNPs in the polymer host matrix this scattering is rather weak. 
Results of our calculations did not reveal any significant contribution to $\langle \widetilde{C}_{n}(q) \rangle_{M}$ in the corresponding range of $\mathbf{ q}$ and are not presented here.}
in Figs.~\ref{Fig:Data1}(b) and \ref{Fig:Data1}(d).  
For the first sample the Fourier spectrum $\langle \widetilde{C}_{n}(q) \rangle_{M}$
[Fig.~\ref{Fig:Data1}(b)] has three local maxima for the Fourier components of the orders $n=2,4$ and $6$.
These maxima are located at the same $q$ values as $(200)$, $(300)$ and $(020)$ peaks on the average intensity curve $\langle \langle I(q,\varphi) \rangle_{\varphi}\rangle_{M}$ [Fig.~\ref{Fig:Data1}(a)].
The presence of higher harmonics of $n=2$ (up to $n=6$) indicates a high degree of orientational order 
(similar to bond-orientational order parameters observed in hexatic phase of a liquid crystal \cite{Kurta4}). 
For the second sample the Fourier spectrum $\langle \widetilde{C}_{n}(q) \rangle_{M}$ has a strong maximum only at the position corresponding to $(020)$ ring.
In this case, the strongest contribution is given by the Fourier component\footnote{The Fourier component of the order $n=1$ also gives a non-zero contribution to the spectra in Figs.~\ref{Fig:Data1}(b) and \ref{Fig:Data1}(d). 
The origin of this contribution will be clarified in the future work.} $n=2$, whereas $n=4$ is significantly weaker, and $n=6$ is about noise level. 
These results indicate a coupling of the orientational anisotropy of different interplanar distances associated with $\pi-\pi$ stacking and alkyl side chains.
Interestingly, for the mixed orientation of domains the same Fourier components (with $n=2,4$ and $6$) contribute to Fourier spectra of the CCF at $\mathbf{q}$ corresponding to all $(h00)$
and $(020)$ peaks [see Fig.~\ref{Fig:Data1}(b)]. One can also see a slight difference of the shape for the $(020)$ radial intensity peak and wells as for the most
intense $n = 2$ Fourier component for two samples [see Figs.~\ref{Fig:Data1}]. A possible reason for the broadening of the $(020)$ peak in the case of mixed orientation is due to a contribution of the higher order $(400)$
peak and P3HT backbone related planes. 

\section{Conclusions}
We applied the XCCA for the analysis of x-ray scattering data from two P3HT films with addition of gold nanoparticles.
Despite similar fabrication protocol for both films, probing with nanobeam on the local scale reveals structural difference between two systems.
While the radial intensity curves look very similar for two samples, 
the Fourier spectra of the intensity cross-correlation functions clearly indicate different degree of structural order in the systems. 
For the sample with dominating edge-on orientation of domains main contribution is given by $n=2,4$ components.  
Interestingly, for the film with mixed orientation an additional contribution to the Fourier spectra with $n=6$ is clearly observable. 
The results suggest a coupling of the orientational anisotropy with in-plane ordering of $\pi-\pi$ conjugated network that requires further investigation.

\ack
We are thankful to the group of T. Salditt, especially to M. Osterhoff and S. Kalbfleisch, for providing support of G{\"o}ttingen instrument for nano-imaging with x-rays (GINI-X), 
and also A. Zozulya for help during the experiment. We acknowledge fruitful discussions and support of this project by E. Weckert.
We thank J. A. Sellberg and T. J. Lane for discussions concerning correlation functions, and S. Funari for a careful reading of the paper.
We gratefully acknowledge Maria Vittoria Russo, Rome and Ullrich Pietsch, Siegen for helpful discussions.
Part of this work was supported by BMBF Proposal 05K10CHG `'Coherent Diffraction Imaging and Scattering of Ultrashort Coherent Pulses with Matter`'
in the framework of the German-Russian collaboration `'Development and Use of Accelerator-Based Photon Sources`' and the Virtual Institute VH-VI-403 of the
Helmholtz Association. This work was also partially supported by the Department of Chemistry, Sapienza University of Rome through the Supporting Research Initiative 2013 and BMBF (project Nr 05K10PSC).

\section*{References}
\bibliography{XCCA_ConjNetw}

\providecommand{\newblock}{}
\begin{thebibliography}{10}
\expandafter\ifx\csname url\endcsname\relax
  \def\url#1{{\tt #1}}\fi
\expandafter\ifx\csname urlprefix\endcsname\relax\def\urlprefix{URL }\fi
\providecommand{\eprint}[2][]{\url{#2}}
% Bibliography created with iopart-num v2.0
% /biblio/bibtex/contrib/iopart-num

\bibitem{Salleo}
Salleo A, Kline R~J, Delongchamp D~M and Chabinyc M~L 2010 {\em Adv. Mater.\/}
  {\bf 22} 3812--3838

\bibitem{Prosa}
Prosa T~J, Winokur M~J, Moulton J, Smith P and Heeger A~J 1992 {\em
  Macromolecules\/} {\bf 25} 4364--4372

\bibitem{Rahimi}
Rahimi K, Botiz I, Stingelin N, Kayunkid N, Sommer M, Koch F~P, Nguyen H,
  Coulembier O, Dubois P, Brinkmann M and Reiter G 2012 {\em Angew. Chem. Int.
  Ed. Engl.\/} {\bf 51} 11131--11135

\bibitem{Kim}
Kim D~H, Jang Y, Park Y~D and Cho K 2006 {\em Macromolecules\/} {\bf 39}
  5843--5847

\bibitem{Sirring}
Sirringhaus H and et~al 1999 {\em Nature\/} {\bf 401} 685--688

\bibitem{Shabi}
Shabi T~S, Mikayelyan E, Grigorian S, Pietsch U, Koenen N, Scherf U, Kayunkid N
  and Brinkmann M 2012 {\em Macromolecules\/} {\bf 45} 5575--5585

\bibitem{Wochner1}
Wochner P, Gutt C, Autenrieth T, Demmer T, Bugaev V, Diaz-Ortiz A, Duri A,
  Zontone F, Gr{\"u}bel G and Dosch H 2009 {\em Proc. Nat. Acad. Sci.\/} {\bf
  106} 11511--11514

\bibitem{Altarelli}
Altarelli M, Kurta R~P and Vartanyants I~A 2010 {\em Phys. Rev. B\/} {\bf 82}
  104207

\bibitem{AltarelliErr}
Altarelli M, Kurta R~P and Vartanyants I~A 2012 {\em Phys. Rev. B\/} {\bf 86}
  179904(E)

\bibitem{Kurta1}
Kurta R~P, Altarelli M, Weckert E and Vartanyants I~A 2012 {\em Phys. Rev. B\/}
  {\bf 85} 184204

\bibitem{Kurta2}
Kurta R~P, Dronyak R, Altarelli M, Weckert E and Vartanyants I~A 2013 {\em New
  J. Phys.\/} {\bf 15} 013059

\bibitem{Kurta3}
Kurta R~P, Altarelli M and Vartanyants I~A 2013 {\em Adv. Cond. Matt. Phys.\/}
  {\bf 2013} 959835

\bibitem{Gutt}
Gutt C, Grodd L, Mikayelyan E, Pietsch U, Kline R~J and Grigorian S 2013 {\em
  Sci. Rep.\/}  (submitted)

\bibitem{Kurta4}
Kurta R~P, Ostrovskii B~I, Singer A, Gorobtsov O~Y, Shabalin A, Dzhigaev D,
  Yefanov O~M, Zozulya A~V, Sprung M and Vartanyants I~A 2013 {\em Phys. Rev.
  E\/} {\bf 88} 044501

\bibitem{Kalbfleisch}
Kalbfleisch S, Neubauer H, Kruger S~P, Bartels M, Osterhoff M, Mai D~D,
  Giewekemeyer K, Hartmann B, Sprung M and Salditt T 2011 {\em AIP Conf.
  Proc.\/} {\bf 1365} 96--99

\bibitem{Fratoddi}
Fratoddi I, Venditti I, Battocchio C, Polzonetti G, Bondino F, Malvestuto M,
  Piscopiello E, Tapfer L and Russo M~V 2011 {\em J. Phys. Chem. C\/} {\bf 115}
  15198–15204

\bibitem{Matassa}
Matassa R, Fratoddi I, Rossi M, Battocchio C, Caminiti R and Russo M~V 2012
  {\em J. Phys. Chem. C\/} {\bf 116} 15795--15800

\bibitem{Quintiliani}
Quintiliani M, Bassetti M, Pasquini C, Battocchio C, Fratoddi I, Rossi M, Mura
  F, Matassa R and Russo M~V   (in preparation)

\end{thebibliography}

\end{document}